\documentclass[12pt]{elsart}
\usepackage{feynmp,epsfig,graphics}
\newcommand{\MZ}{M^{\!\!\!\!\!^o}}
\newcommand{\tr}{\rm tr \,}

\begin{document}
GSI-Preprint-2003-29
\begin{frontmatter}
\title{Open-charm meson resonances with negative strangeness}

\author[GSI]{J. Hofmann}
\author[GSI,TU]{and M.F.M. Lutz}
\address[GSI]{Gesellschaft f\"ur Schwerionenforschung (GSI)\\
Planck Str. 1, 64291 Darmstadt, Germany}
\address[TU]{Institut f\"ur Kernphysik, TU Darmstadt\\
D-64289 Darmstadt, Germany}
\begin{abstract}
We study heavy-light meson resonances with quantum numbers $J^P=0^+$ and $J^P=1^+$ in
terms of the non-linear chiral SU(3) Lagrangian. Adjusting the free parameters that arise
at subleading order to reproduce the mass of the $D(2420)$ resonance as well as the
new states established recently by the BABAR, CLEO  and BELLE collaborations we obtain refined
masses for the anti-triplet and sextet states. Bound states of antikaons at the
$D(1867)$ and $D(2008)$ mesons are predicted at 2352 MeV ($J^P=0^+$) and 2416 MeV ($J^P=1^+$).
In addition we anticipate a narrow scalar state of mass 2389 MeV with $(I,S)=(\frac{1}{2},0)$.
\end{abstract}
\end{frontmatter}

\section{Introduction}

In  a recent work \cite{KL0307133} it was demonstrated that chiral SU(3) symmetry predicts
parameter-free heavy-light $J^P\!=\!0^+$ and $J^P\!=\!1^+$ meson resonances. The states
form an anti-triplet and a sextet representation of the SU(3) flavor group. In the
open-charm sector the recently discovered narrow state of mass 2.317 GeV \cite{BaBar} and
2.463 GeV \cite{CLEO} were recovered as part of the strongly bound anti-triplet
states. The missing $(\frac{1}{2},0)$ triplet states were later announced by the
BELLE collaboration \cite{BELLE} finding broad resonance structures at 2.308 GeV  and 2.427 GeV
in the $J^P\!=\!0^+$ and $J^P\!=\!1^+$  channels respectively. Such states were first
predicted in \cite{NRZ93,BH94} based on the spontaneous breaking of chiral symmetry and
heavy-quark symmetry.
The spectacular experimental discoveries triggered a flurry of theoretical publications
\cite{NRZ93,BH94,BEH03,NRZ03,all-papers}.

The purpose of this Letter is to study the properties of the newly predicted sextet
states \cite{KL0307133} in  the open charm sector as chiral correction terms are
considered. Since contrary to the open bottom sector in which already the leading
order computation \cite{KL0307133} predicts weakly bound $\bar K \,B$ isospin-zero states
the binding in the $\bar K\,D$ system is not quite sufficiently strong to form a bound
state at leading order. In \cite{KL0307133} it was pointed out that chiral correction
terms should provide additional attraction to form open charm bound states with negative
strangeness also. The argument was based on the identification of the $D(2420)$ resonance
as a member of a SU(3)-sextet. Our opinion differs here from the traditional approach
\cite{IW91,KKP92,C95,Casalbuoni} which interprets the latter state as part of an anti-triplet.
Since the leading order computation underestimates the binding
of the latter resonance by about 130 MeV correction terms are expected to provide additional
binding for the sextet states in particular in the strangeness minus one sector.

In this work we apply the $\chi$-BS(3) approach developed
originally for meson-baryon scattering \cite{LK00,LK01,LH02,LK02,Granada,Copenhagen} but
recently also applied successfully to meson-meson scattering \cite{KL0307133,LK03}.
For earlier works on meson-meson scattering based on various schemes see
\cite{Rupp86,WI90,JPHS95,OOP99,NVA02,NP02}. Using the
chiral SU(3) Lagrangian involving heavy-light $J^P\!=\!0^-$ and $J^P\!=\!1^-$ fields that
transform non-linear under the chiral SU(3) group a coupled-channel computation of the
meson-meson scattering amplitude in the open-charm sector is performed (see
also \cite{Flyn-Nieves-91}). Leading and subleading terms are considered in the chiral
expansion of the interaction kernel. At subleading order three unknown parameters arise
that can be adjusted to reproduce accurately the scalar or axial-vector spectrum. The
values for the parameters determined independently in the scalar and axial-vector sector are
reasonably close consistent with the expectation of heavy-quark symmetry. The central result
of this work is the prediction of bound states of antikaons at the $D(1867)$ and $D(2008)$ mesons
of mass 2352 MeV ($J^P=0^+$) and 2416 MeV ($J^P=1^+$). In addition we anticipate a narrow scalar
state of mass 2389 MeV with $(I,S)=(\frac{1}{2},0)$.

\section{The $\chi$-BS(3) approach}

The starting point to study the scattering of Goldstone bosons off
heavy-light mesons is the chiral SU(3) Lagrangian.
We identify the leading-orders Lagrangian
density \cite{Wein-Tomo,Wise92,YCCLLY92,BD92,Jenkins94} describing the s-wave interaction of
Goldstone bosons with pseudo-scalar and vector-mesons. In order to exploit the heavy-quark
symmetry it is convenient to introduce a Dirac valued field \cite{Wise92,YCCLLY92,BD92,Jenkins94,Casalbuoni}
\begin{eqnarray}
H(x)= \frac{1}{2}\,\big( P_\mu(x)\,\gamma^\mu - \gamma_5\,P(x) \big)\,,\qquad
\bar H(x) = \gamma_0\,H^\dagger(x) \,\gamma_0\, ,
\label{}
\end{eqnarray}
which encodes the massive pseudo-scalar field, $P(x)$, and the vector-meson field, $P_\mu(x)$.
We work directly with the relativistic chiral Lagrangian so that the field $H(x)$
carries dimension one and does not involve the 4-velocity $v_\mu $ of the heavy-meson.
Neglecting terms that are suppressed in the heavy-quark mass limit we
write
\begin{eqnarray}
{\mathcal L}(x) &=& \frac{1}{2}\,\tr \Big( \bar H(x) \, \big(
\partial^2 + {\MZ}^{_{\;2}}  + 4\,c_0\,(\tr \chi_0 )- 4\,c_1 \,\chi_0\big) \,H(x) \Big)
\nonumber\\
&+& \frac{1}{8\,f^2}\,\tr \Big( \Big[
(\partial^\nu H(x))\,\bar H(x)
-H(x)\, (\partial^\nu \bar H(x) ) \Big]
\,\Big[\Phi (x) , (\partial_\nu\,\Phi(x))\Big]_- \Big)
\nonumber\\
&+& \frac{g\,\MZ}{f}\,\tr \Big(\bar H(x)\,\gamma_5\,\gamma^\mu\,(\partial_\mu \Phi(x))\,H(x) \Big)
\nonumber\\
&-&\frac{c_0}{f^2}\,\tr \Big( \Phi(x) \,\chi_0 \,\Phi(x) \Big)\,
\tr \Big( \bar H(x)  \, H(x) \Big)
\nonumber\\
&+& \frac{c_1}{4\,f^2}\,\tr \Big( \bar H(x) \, \Big[ \Phi(x), \Big[ \Phi(x) ,\chi_0 \Big]_+\Big]_+ \,H(x) \Big)
\nonumber\\
&-& \frac{c_2}{f^2}\, \tr \Big( \bar H(x)\, H(x) \Big)
\,\tr \Big( (\partial^\mu\,\Phi(x))\,(\partial_\mu\,\Phi(x)) \Big)
\nonumber\\
&-& \frac{c_3}{f^2}\, \tr \Big( \bar H(x)
\,  (\partial^\mu\,\Phi(x))\,(\partial_\mu\,\Phi(x)) \,H(x) \Big)\,,
\nonumber\\
\chi_0 &=&\frac{1}{3} \left( m_\pi^2+2\,m_K^2 \right)\,1
+\frac{2}{\sqrt{3}}\,\left(m_\pi^2-m_K^2\right) \lambda_8 \,,
\label{WT-term}
\end{eqnarray}
where $\Phi$ is the Goldstone bosons field.
The parameter $f$ in (\ref{WT-term}) is known from the weak decay process of the
pions. We use $f= 90$ MeV through out this work. The parameter
$g$ can be adjusted the partial decay width $D_+(2008)\to \pi^+\,D(1867)$. Using
the latest values \cite{PDG02} of $64 \pm 15 $ keV one obtains $g\,\MZ \simeq 1155 \pm 135 $ MeV
at tree-level. The symmetry breaking parameter $c_1$ can be determined by the mass splitting
of the open charm pseudo-scalar or vector-meson ground states,
\begin{eqnarray}
M_{D^{(s)}_0}^2 -M_{D_0}^2 = 4\,c_1 \,(m_K^2-m_\pi^2)  \,,\qquad
M_{D^{(s)}_1}^2 -M_{D_1}^2 = 4\,c_1 \,(m_K^2-m_\pi^2) \,,
\label{}
\end{eqnarray}
which leads to $c_1 \simeq 0.44$ and $c_1 \simeq 0.47 $ for the pseudo-scalar and
vector states respectively. We use the averaged value $c_1 \simeq 0.45$ in this work.
The remaining three dimension less parameters $c_0,c_2,c_3$ will be determined in
this work. Formally the latter parameters scale with $c_i \sim \MZ $.

Since we will assume perfect isospin symmetry it is convenient to decompose
the fields into their isospin multiplets. The fields can be written in terms of
isospin multiplet fields like $K =(K^{(+)},K^{(0)})^t $ and $D=(D^{(+)},D^{(0)})^t$,
\begin{eqnarray}
&& \Phi = \tau \cdot \pi (140)
+ \alpha^\dagger \cdot  K (494) +  K^\dagger(494) \cdot \alpha
+ \eta(547)\,\lambda_8\,,
\nonumber\\
&& P_{\phantom{\mu}}  = {\textstyle{1\over \sqrt{2}}}\,\alpha^\dagger \cdot D_{\phantom{\mu}}(1867)
- {\textstyle{1\over \sqrt{2}}}\,D^t_{\phantom{\mu}}(1867)\cdot \alpha   +  i\,\tau_2\,D_{\phantom{\mu}}^{(s)}(1969) \,,
\nonumber\\
&& P_\mu = {\textstyle{1\over \sqrt{2}}}\,\alpha^\dagger \cdot D_\mu(2008)
- {\textstyle{1\over \sqrt{2}}}\,D^t_\mu(2008)\cdot \alpha   +  i\,\tau_2\,D^{(s)}_{\mu}(2110) \,,
\nonumber\\
&& \alpha^\dagger = {\textstyle{1\over \sqrt{2}}}\,(\lambda_4+i\,\lambda_5 ,\lambda_6+i\,\lambda_7 )\,,\qquad
\tau = (\lambda_1,\lambda_2, \lambda_3)\,,
\label{field-decom}
\end{eqnarray}
where the matrices $\lambda_i$ are the standard Gell-Mann generators of the SU(3) algebra.
The numbers in the brackets recall the approximate masses of the fields in units of MeV.
Even though we did not write down the relevant term in (\ref{WT-term}) describing the
mass splitting of the $0^-$ and $1^-$ states we will use physical masses as
given in (\ref{field-decom}) through out this work.

\begin{table}
\tabcolsep=-.1mm
\begin{tabular}{||cccccccccccc||}
\hline\hline
\multicolumn{3}{||c|}{$(\frac{1}{2},+2)$} &
\multicolumn{3}{c|}{$(0,+1)$} &
\multicolumn{3}{c|}{$(1,+1)$} &
\multicolumn{3}{c|}{$(\frac12,0)$}
\\ \hline
\multicolumn{3}{||c|}{$(D_s\,K)$} &
\multicolumn{3}{c|}{$\left(\begin{array}{c} ({\textstyle{1\over
\sqrt{2}}}\,D^t \, i\,\sigma_2\, K)\\ (D_s\,\eta )
\end{array}\right)$}&
\multicolumn{3}{c|}{$\left(\begin{array}{c} (D_s \, \pi)\\ (
{\textstyle{1\over
\sqrt{2}}}\,D^t \, i\,\sigma_2\,\sigma\, K)
\end{array}\right)$} &
\multicolumn{3}{c|}{$\left(\begin{array}{c} ({\textstyle{1\over
\sqrt{3}}}\,\pi \cdot
\sigma\, D)\\ ( \eta \,D) \\ (D_s\,i\,\sigma_2\,\bar K^t)
\end{array}\right)$}
\\\hline\hline
\multicolumn{4}{||c|}{\phantom{xxxxxx}$(\frac32,0)$\phantom{xxxxxx}} &
\multicolumn{4}{c|}{\phantom{xxxxxx}$(0,-1)$\phantom{xxxxxx}} &
\multicolumn{4}{c|}{$(1,-1)$}
\\\hline
\multicolumn{4}{||c|}{$( \pi \cdot T\,D)$} &
\multicolumn{4}{c|}{$( {\textstyle{1\over \sqrt{2}}}\,\bar K
\,D)$} &
\multicolumn{4}{c|}{$({\textstyle{1\over \sqrt{2}}}\,\bar K
\,\sigma\,D)$}
\\\hline\hline
\end{tabular}
\caption{Coupled-channel states with $(I,S)$.}
\label{tab:states}
\end{table}

The scattering problem decouples into seven orthogonal
sectors specified by isospin ($I$) and strangeness ($S$) quantum numbers,
\begin{eqnarray}
(I,S)= (({\textstyle{1\over{2}}},+2), (0,+1), (1, +1), ({\textstyle{1\over{2}}}, 0),
({\textstyle{3\over{2}}}, 0), (0,-1), (1,-1))\,.
\label{lwt}
\end{eqnarray}
In Tab. \ref{tab:states} the channels that contribute in a given sector $(I,S)$ are listed.
Heavy-light meson resonances with quantum numbers $J^P\!=\!0^+$ and $J^P\!=\!1^+$ manifest
themselves as poles in the s-wave scattering amplitudes, $M^{(I,S)}_{J^P}(\sqrt{s}\,)$, which
in the $\chi-$BS(3) approach \cite{LK02,LK03} take the simple form
\begin{eqnarray}
&&  M^{(I,S)}_{J^P}(\sqrt{s}\,) = \Big[ 1- V^{(I,S)}_{J^P}(\sqrt{s}\,)\,J^{(I,S)}_{J^P}(\sqrt{s}\,)\Big]^{-1}\,
V^{(I,S)}(\sqrt{s}\,)\,.
\label{final-t}
\end{eqnarray}
We first specify the contributions to the
effective interaction kernel $V^{(I,S)}_{J^P}(\sqrt{s}\,)$ in (\ref{final-t}) from  the
Weinberg-Tomozawa term and the subleading interaction terms proportional to the
parameters $c_i$ as introduced in (\ref{WT-term}),
\begin{eqnarray}
&& V^{(I,S)}(\sqrt{s}\,) = \frac{C^{(I,S)}}{8\,f^2}\, \Big(
3\,s-M^2-\bar M^2-m^2-\bar m^2
\nonumber\\
&& \qquad \qquad \qquad \qquad \qquad -\frac{M^2-m^2}{s}\,(\bar M^2-\bar m^2)\Big)
\nonumber\\
&& \qquad  +\, 2\,\frac{m_\pi^2}{f^2}\,\Big( c_0\,C^{(I,S)}_{\pi,0}+c_1\,C^{(I,S)}_{\pi,1}\Big)
+ 2\,\frac{m_K^2}{f^2}\,\Big( c_0\,C^{(I,S)}_{K,0}+c_1\,C^{(I,S)}_{K,1}\Big)
\nonumber\\
&& \qquad  +\,
\Big( c_2\,\frac{C^{(I,S)}_2}{s\,f^2} +c_3\,\frac{C^{(I,S)}_3}{s\,f^2} \Big)
\,\Big(s-\bar M^2+\bar m^2\Big)\,\Big(s-M^2+m^2\Big) \,,
\label{VWT}
\end{eqnarray}
where $(m,M)$ and $(\bar m, \bar M)$ are the masses of initial and final mesons. We use
capital $M$ for the masses of heavy-light mesons and small $m$ for the masses of the Goldstone
bosons. The s-wave interaction kernels are identical for the two scattering problems considered
here. Following \cite{LK03} we neglect here the mixing of the s- with a d-wave channel in the
$1^+$ sector. Such effects are largely suppressed. We continue with a presentation of the
contributions of the s- and u-channel exchange terms
that are proportional to $g^2$. In this case one has to discriminate the $0^+$ and
$1^+$ channels,
\begin{eqnarray}
&& V^{(I,S)}_{0^+}(\sqrt{s}\,) =  g^2\,\frac{C^{(I,S)}_{s-ch}}{4\,s\,f^2}\,
 \Big(s-\bar M^2+\bar m^2\Big)\,\Big(s-M^2+m^2\Big)
\nonumber\\
&& \qquad \qquad +\,g^2\, \frac{C^{(I,S)}_{u-ch}}{f^2}\,\int_{-1}^1 \frac{d x}{2}
\frac{(\bar q\cdot q) \,\mu^2- (\bar m^2 - \bar q\cdot p )\,
(m^2- \bar p\cdot q )}{M^2+\bar M^2 -\mu^2 -s +2\,\bar q\cdot q} \,,
\nonumber\\
&& V^{(I,S)}_{1^+}(\sqrt{s}\,) = -g^2\, \frac{C^{(I,S)}_{u-ch}}{2\,f^2}\,\int_{-1}^1 \frac{d x}{2}
\frac{\mu^2\,\bar p_{cm}\,p_{cm}\,x\,(1-x^2)}{M^2+\bar M^2 -\mu^2 -s +2\,\bar q\cdot q} \,,
\nonumber\\
&& \bar q \cdot q = \sqrt{\bar m^2+\bar p_{\rm cm}^2}\,\sqrt{m^2+p_{\rm cm}^2}
-\bar p_{\rm cm}\,p_{\rm cm}\,x \,,
\nonumber\\
&& \bar q\cdot p = -\bar q\cdot q  + \frac{s-\bar M^2+\bar m^2}{2}\,,
\qquad \bar p \cdot q = -\bar q\cdot q  + \frac{s- M^2+m^2}{2}\,,
\nonumber\\
&& \sqrt{s}= \sqrt{M^2+p_{cm}^2}+ \sqrt{m^2+p_{cm}^2}=
\sqrt{\bar M^2+\bar p_{cm}^2}+ \sqrt{\bar m^2+\bar p_{cm}^2} \,,
\label{s-u-channel}
\end{eqnarray}
where $\mu$ is the appropriate mass of the heavy-meson exchanged in the u-channel.
Note that we identified $\MZ \simeq \mu$ in (\ref{s-u-channel}).
As is evident from the representation (\ref{s-u-channel}) the contribution of the s- and
u-channel terms scale with $\sim (\MZ\,)^0$ as compared to the linear scaling $\sim (\MZ\,)^1$ of
the terms in (\ref{VWT}), which follows from $c_i \sim \MZ $.
Therefore we expect the contributions (\ref{s-u-channel}) to be of
subsubleading importance.
The matrix of coefficients $C^{(I,S)}$ that characterize the interaction strength in
any given channel are presented in Tab. \ref{tab:coeff}.

We turn to the loop functions introduced in (\ref{final-t}). The latter
are diagonal in the coupled-channel space and depend on
whether to scatter Goldstone bosons off pseudo-scalar or vector heavy-light mesons,
\begin{eqnarray}
&& J_{0^+}(\sqrt{s}\,) = I(\sqrt{s}\,)-I(\mu_{0^+}^{(I,S)})\,,
\nonumber\\
&& J_{1^+}(\sqrt{s}\,) =
\Big(1 + \frac{p_{\rm cm}^2}{3\,M^2} \Big)\,
\Big(I(\sqrt{s}\,)-I(\mu_{1^+}^{(I,S)}) \Big)\,,
\nonumber\\
\nonumber\\
&& I(\sqrt{s}\,)=\frac{1}{16\,\pi^2}
\left( \frac{p_{cm}}{\sqrt{s}}\,
\left( \ln \left(1-\frac{s-2\,p_{cm}\,\sqrt{s}}{m^2+M^2} \right)
-\ln \left(1-\frac{s+2\,p_{cm}\sqrt{s}}{m^2+M^2} \right)\right)
\right.
\nonumber\\
&&\qquad \qquad + \left.
\left(\frac{1}{2}\,\frac{m^2+M^2}{m^2-M^2}
-\frac{m^2-M^2}{2\,s}
\right)
\,\ln \left( \frac{m^2}{M^2}\right) +1 \right)+I(0)\;.
\label{i-def}
\end{eqnarray}
As expected from heavy-quark symmetry the two loop functions $J_{0^+}(\sqrt{s}\,)$ and
$J_{1^+}(\sqrt{s}\,)$ in (\ref{i-def}) differ by a term suppressed with $1/M^2$ only.
A crucial ingredient of the $\chi-$BS(3) scheme is its approximate crossing symmetry guaranteed
by a proper choice of the subtraction scale $\mu_{J^P}^{(I,S)}$. Referring to the detailed discussions
in \cite{LK02,Granada,Copenhagen,LK03} we obtain
\begin{eqnarray}
&& \mu_{0^+}^{(I,0)} =  M_{D(1867)}\,,\quad
\mu_{0^+}^{(I, \pm 1)} = M_{D_s(1969)}\,,\quad
\mu_{0^+}^{(I, 2)} = M_{D(1867)} \,,
\nonumber\\
&& \mu_{1^+}^{(I,0)} =  M_{D(2008)}\,,\quad
\mu_{1^+}^{(I, \pm 1)} = M_{D_s(2110)}\,,\quad
\mu_{1^+}^{(I, 2)} = M_{D(2008)} \,.
\label{mu-def}
\end{eqnarray}
With (\ref{final-t},\ref{VWT},\ref{s-u-channel},\ref{i-def},\ref{mu-def}) the brief exposition of the $\chi-$BS(3)
approach as  applied to heavy-light meson resonances is completed.

\begin{table}
\tabcolsep=1.0mm
\begin{center}
\begin{tabular}{||c||c||c|c|c|c|c|c|c|c|c||}
\hline
\hline
(I,S) & Channel & $C^{(I,S)}$ & $C^{(I,S)}_{\pi,0}$ & $C^{(I,S)}_{\pi,1}$ & $C^{(I,S)}_{K,0}$ & $C^{(I,S)}_{K,1}$ & $C^{(I,S)}_2$ & $C^{(I,S)}_3$ & $C^{(I,S)}_{s-ch}$ & $C^{(I,S)}_{u-ch}$ \\
\hline
\hline
($\frac12$,+2)  & 11    &$-1$     &0  &0      &4  &$-1$     &2  &$\frac12$  & 0 & 2\\
\hline
\hline
(0,+1)      & 11    &2      &0  &0      &4  & 0     &2  &0     & 4& 0 \\
\hline
        & 12    &$\sqrt3$   &0  &$-\frac{\sqrt3}2$&0 & $\frac5{2\sqrt3}$&0   &$-\frac1{2\sqrt3}$ &$\frac{4}{\sqrt{3}}$& $-\frac{2}{\sqrt{3}}$\\
\hline
        & 22    &0      &$-\frac43$& $-\frac{2}{3}$  & $\frac{16}{3}$  & 0 & 2   &$\frac13$  &$\frac{4}{3}$& $\frac{4}{3}$ \\
\hline
\hline
(1,+1)      & 11    &0      &4  &$-2$     &0  &0      &2  &1     &0 & 0 \\
\hline
        & 12    &1      &0  &$-\frac12$  &0  &$-\frac12$  &0  &$\frac12$  &0& $-2$\\
\hline
        & 22    &0      &0  &0      &4  &$-2$     &2  &1    &0& 0  \\
\hline
\hline
($\frac12$,\,0)   & 11    &2      &4  & $-1$     &0  &0      &2  &$\frac12$  & 3& $-1$\\
\hline
        & 12    &0      &0  & $-1$      &0  &0      &0  &$\frac12$  & $-1$& $-1$\\
\hline
        & 13    &0      &0  &$\sqrt{\frac38}$&0    & $\sqrt{\frac38}$&0    &$-\sqrt{\frac38}$ &$\sqrt{6}$& 0\\
\hline
        & 22    &$\sqrt{\frac32}$& $-\frac43$& 1  &$\frac{16}3$& $-\frac83$&2  &$\frac56$ & $\frac{1}{3}$& $\frac{1}{3}$\\
\hline
        & 23    & $-\sqrt{\frac32}$&0    &$-\sqrt{\frac38}$&0 & $\frac5{2\sqrt6}$&0   &$-\frac1{2\sqrt6}$ & $-\sqrt{\frac{2}{3}}$& $\sqrt{\frac{8}{3}}$\\
\hline
        & 33    &1      &0  &0      &4  & $-1$   &2  &$\frac12$  & 2& 0\\
\hline
\hline
($\frac32$,\,0)   & 11    &$-1$     &4  &$-1$     &0  &0      &2  &$\frac12$ & 0& 2 \\
\hline
\hline
(0,-1)      & 11    &1      &0  &0      &4  & $-3$     &2  &$\frac32$ & 0& $-2$\\
\hline
\hline
(1,-1)      & 11    &$-1$     &0  &0      &4  & $-1$     &2  &$\frac12$ & 0& 2 \\
\hline
\hline
\end{tabular}
\caption{The coefficients $C^{(I,S)}$ that characterize the  interaction of Goldstone bosons with
heavy-meson fields $H$ as introduced in (\ref{VWT},\ref{s-u-channel}) for
given isospin (I) and strangeness (S). The channel ordering is specified in
Tab. \ref{tab:states}.}
\label{tab:coeff}
\end{center}
\end{table}

\section{Results}

\begin{table}
\begin{center}
\begin{tabular}{|c||c|c|c|c||c|c|c|c|}
\hline
J$^P_{(I,S)}$ &$0^+_{(0,1)}$&$0^+_{(\frac12,0)}$&$0^+_{(\frac12,0)}$&$0^+_{(0,-1)}$&  $1^+_{(0,1)}$&$1^+_{(\frac12,0)}$&$1^+_{(\frac12,0)}$&$1^+_{(0,-1)}$\\
\hline
\hline
M$_R$   &2318   &2255   &2389   &2352   &2464   &2300   &2422   &2416\\
\hline
$\Gamma$&-  &360    &0.0    &-  &-  &300    &23 &-  \\
\hline
$|g_1|$ &3.7    &2.1    &0.0    &2.5    &4.2    &1.4    &0.8    &4.6    \\
\hline
$|g_2|$ &3.7    &0.9    &2.3    &-  &5.3    &0.0    &3.8    &-  \\
\hline
$|g_3|$ &-  &2.4    &3.6    &-  &-  & 0.9    &4.1    &-  \\
\hline
\hline
g   &$c_0$  &$c_1$  &$c_2$  &$c_3$  &$c_0$  &$c_1$  &$c_2$  &$c_3$  \\
\hline
0.55    &0.95   &0.45   &-1.64  &1.6    &1.45   &0.45   &-2.13  &2.45   \\
\hline
\end{tabular}
\caption{Coupling constants $c_i$ (see (\ref{WT-term})) for scalar and axial-vector
open-charm mesons. The resonance coupling constants $g_i$ are defined in (\ref{def-g})
with respect to channel labelling of Tab. \ref{tab:states}.}
\label{tab:1}
\end{center}
\end{table}

\begin{figure}[t]
\begin{center}
\includegraphics[clip=true,width=12cm]{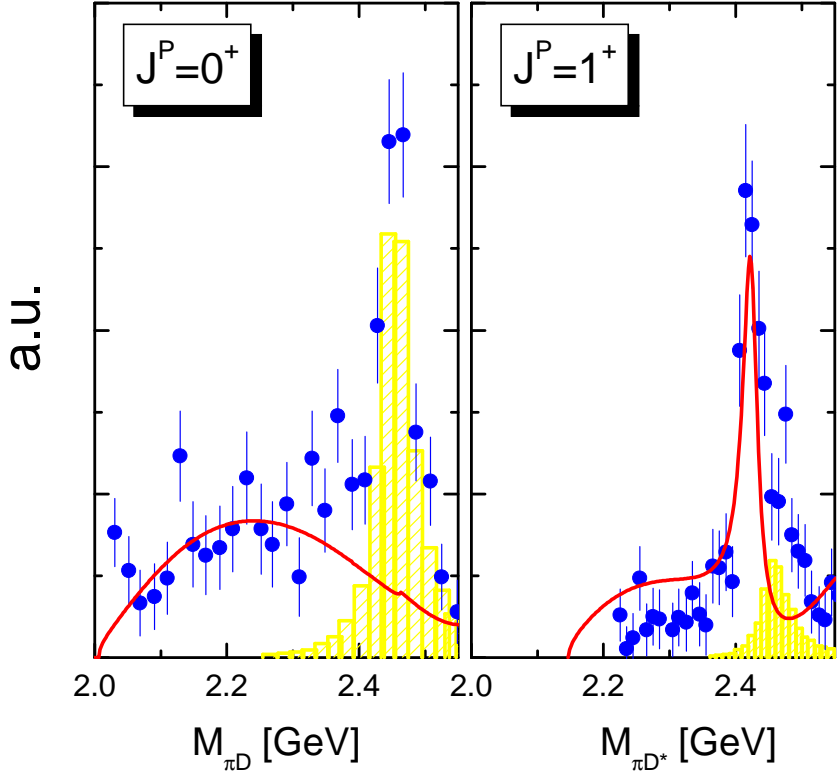}\\
\end{center}
\caption{Mass spectra of the $(\frac{1}{2},0)$-resonances as seen in the
$ \pi \,D(1867)$-channel (l.h. panel $J^P=0^+$) and $ \pi \,D(2008)$-channel (r.h. panel $J^P=1^+$).
The solid lines show the theoretical mass distributions. The data are taken from \cite{BELLE}
as obtained from the $B \to \pi \,D(1867)$  and $B \to  \pi \,D(2008)$
decays. The histograms indicate the contribution from the
$J=2$ resonances $D(2460)$ as given in \cite{BELLE}.}\label{fig:1}
\end{figure}

\begin{figure}[t]
\begin{center}
\includegraphics[clip=true,width=12cm]{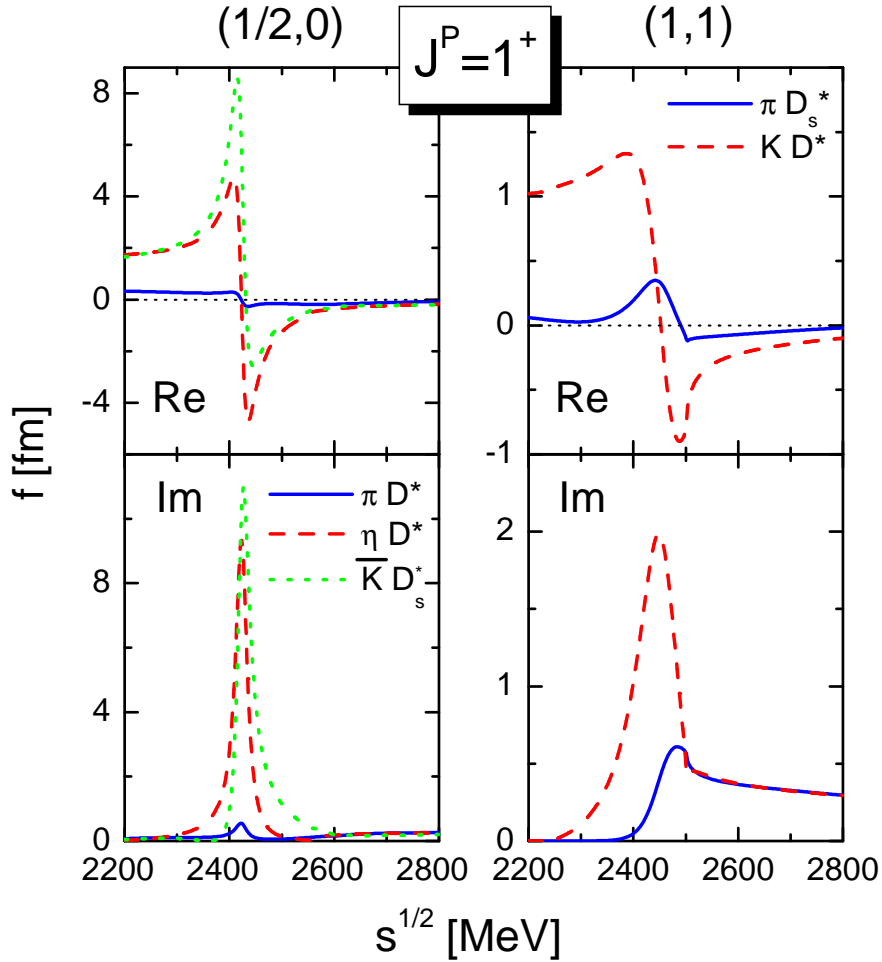}
\end{center}
\caption{Open charm resonances with $J^P =1^+$  and $(I,S)=(\frac{1}{2},0),(0,1)$ as seen in
the scattering of Goldstone bosons of $D(2008)$- and $D_s(2110)$-mesons. Shown are real and imaginary
parts of reduced scattering amplitude, $f_{aa}(\sqrt{s}\,)= M_{aa}(\sqrt{s}\,)/(8\pi \sqrt{s})$.
}\label{fig:2}
\end{figure}

We begin with a presentation of results for the open-charm axial-vector mesons.
At leading order \cite{KL0307133} with $c_i=0$ and $g=0$ using
physical masses for the intermediate states a narrow $(0,1)$-state at 2440 MeV
is generated by coupled-channel dynamics. Furthermore in the $(\frac{1}{2},0)$-sector
a narrow and a broad state at 2552 MeV and 2300 MeV are predicted. The leading
order results are already rather close to the empirical values of
$2463 \pm 2$ MeV \cite{CLEO} for the $(0,1)$-state and
$2421.1 \pm 2.7$ MeV and $2427 \pm 61$ MeV with widths $23.7\pm 6.9$ MeV and
$384 \pm 94$ MeV \cite{BELLE} for the $(\frac{1}{2},0)$-states.
It is straight forward to find the optimal values for the three unknown parameters
$c_{0},c_{2}$ and $c_3$ that arise at subleading order as to reproduce the
empirical pattern quite accurately. In Tab. \ref{tab:1} the resulting parameters
are given. To be precise we use the averaged mass $\mu = 1918$ MeV in (\ref{s-u-channel}).
The typical size of these parameters appears consistent with the
naturalness assumption $c_i \sim \MZ /m_\rho $ suggesting that the chiral expansion is
well converging. The results do not sensitively depend on the value of $g$ within
the range $0.4 < g < 0.7$. The empirical masses of all three states can be reproduced
within experimental errors. The chiral corrections terms pull the narrow
$(\frac{1}{2},0)$-state down and the broad $(\frac{1}{2},0)$-state up close to their empirical
masses (see Tab. \ref{tab:1}). In Fig. \ref{fig:1} we confront the spectral distribution
of the $\pi\,D(2008)$-channel measured recently by the BELLE collaboration \cite{BELLE}
with the theoretical distribution obtained from the imaginary part of the
$\pi\,D(2008)$-scattering amplitude in the $(\frac{1}{2},0)$-sector. Within empirical errors
we are able to reproduce the shape of the distribution. The small contribution of a
state with $J=2$ shown in the figure by the histograms \cite{BELLE} is not considered
in this work. In Fig. \ref{fig:2} the scattering amplitude of the $(\frac{1}{2},0)$-sector
are shown. The figure demonstrates that the $D(2420)$ resonance couples dominantly
to the closed $\eta \,D(2008)$- and $\bar K \,D_s(2110)$-channels which are the driving channels
for the dynamic generation of the latter resonance.
Having adjusted all parameters at subleading order we can now turn to the
$(0,-1)$- and $(1,1)$-sectors unconstrained yet by data. We find a $(1,1)$-resonance
at 2445 MeV of width 70 MeV (see Fig. \ref{fig:2}) that couples
dominantly to the $K\,D(2008)$-channel with $|g_2| = 3.0$ and a
narrow $(0,-1)$-state at 2416 MeV. As
anticipated in \cite{KL0307133} the chiral correction terms increase the amount of attraction
in the sextet channel predicting the existence of a $\bar K\,D(2008)$-bound state.

We analyze the properties of the anti-triplet and sextet states in more detail by extracting
coupling constants. If in a given coupled-channel scattering amplitude
$M_{ab}(\sqrt{s}\,) $ a bound or resonance state of mass $M_R$ is present
we determine the coupling constants $g_a$ of that state to the channel $a$ by the condition
\begin{eqnarray}
M_{ab}(\sqrt{s}\,) \simeq - \frac{g^*_a\,g_b\,M_R}{\sqrt{s}-M_R+i\,\Gamma/2} \,,
\label{def-g}
\end{eqnarray}
close to the pole. The resulting parameters are collected in Tab. \ref{tab:1}.
Whereas the chiral correction terms affect the mass of the $(0,1)$-state by 24 MeV only,
the change in the coupling constant is more important. At leading order with
$c_i=0$ and $g=0$ one finds lower values of $|g_1|= 3.3$ and $|g_2|=2.0$ as compared
to the values  4.2 and 5.3 given in Tab. \ref{tab:1}.

\begin{figure}[t]
\begin{center}
\includegraphics[clip=true,width=12cm]{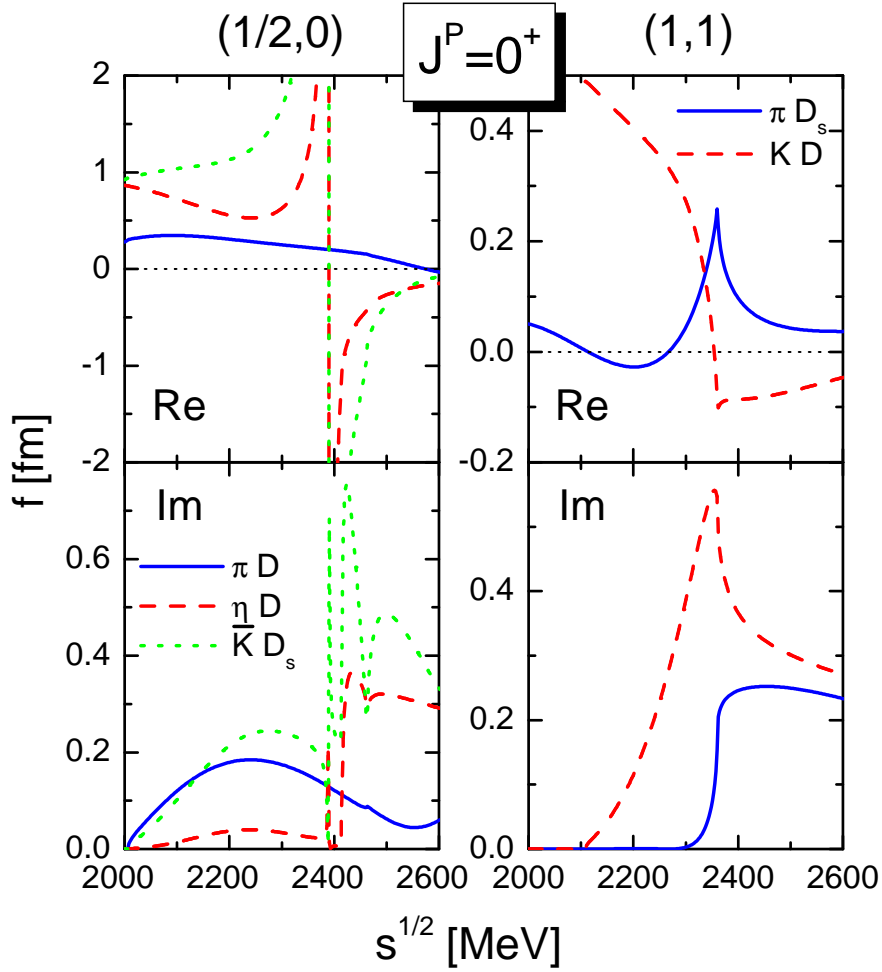}
\end{center}
\caption{Open-charm resonances with $J^P =0^+$  and $(I,S)=(\frac{1}{2},0),(0,1)$ as seen in
the scattering of Goldstone bosons of $D(1867)$- and $D_s(1969)$-mesons. Shown are real and
imaginary parts of reduced scattering amplitude,
$f_{aa}(\sqrt{s}\,)= M_{aa}(\sqrt{s}\,)/(8\pi\sqrt{s})$.}
\label{fig:3}
\end{figure}

We turn to our results for the scalar open-charm mesons. Relying on the heavy-quark symmetry
we could use the values for $c_i$ obtained by reproducing the properties of the axial-vector
mesons. However, since we will make a prediction for the mass of the scalar $(0,1)$-bound state
a more accurate result is expected if we adjust the parameters $c_i$ in the scalar sector.
Therewith subleading interaction terms not displayed in (\ref{WT-term}) which
lead to the independence of the parameters in the scalar and axial-vector sectors are probed.
Of course, consistency requires that the parameters should turn out not too different. We adjust
the parameters to reproduce the narrow $(0,1)$-state at $2317 \pm 3 $ MeV \cite{BaBar} and the
$(\frac{1}{2},0)$-state at $2308\pm 60 $ MeV of width $276 \pm 99$ MeV. The fit is unique since besides
the $(\frac{1}{2},0)$-state we include also the empirical $\pi \,D(1867)$-spectrum obtained
recently by the BELLE collaboration \cite{BELLE}. The fact that the latter spectrum does not
show an additional peak implies that the sextet state couples only very weakly to the $\pi\, D(1867)$-channel. As shown in
Fig. \ref{fig:1} it is possible to adjust the parameters such that the $\pi \,D(1867)$-spectrum
decouples from the sextet state. The theoretical spectrum is again compatible with the
empirical one. Note that the figure includes the contribution of a state with $J=2$ and mass
$2461.6 \pm 5.9$ MeV shown by the dashed histograms \cite{BELLE} but not considered in this work.
Complementary is Fig. \ref{fig:3} which shows the various scattering amplitude in the
$(\frac{1}{2},0)$-sector. The sextet state couples dominantly to the closed
$\eta \,D(1867)$- and $\bar K\,D_s(1969)$-channels. This result is analogous to the corresponding
result for the axial-vector state analyzed in Fig. \ref{fig:2} only that in the scalar sector
the coupling constant of the sextet state to the $\pi \,D(1867)$ is even further suppressed.
The optimal set of parameters obtained are collected in Tab. \ref{tab:1}, which also
includes the resonance parameters. The chiral correction terms push up the
$(\frac{1}{2},0)$-state by 15 MeV to its empirical value. Its coupling constants increase
somewhat as compared to the leading order values $|g_1|=3.3$ and $|g_2|=2.0$.
The values obtained for the coupling constants $c_{0,2,3}$ are reasonably close to the values
obtained in the axial-vector sector. To be precise we should mention that we used the averaged
mass $\mu=2059$ MeV for the s- and u-channel contributions as specified in (\ref{s-u-channel}).

Having fixed all parameters we discuss the predictions for the scalar mesons so far
not observed. The sextet state in the $(\frac{1}{2},0)$-sector at 2389 MeV is quite narrow
with a width of below 1 MeV since its coupling to the $\pi D(1867)$ is much suppressed. We
emphasize that it is difficult to make a precise prediction for the width of that state. If
we only slightly change the set of parameters the sextet state shows up as a narrow peak in the
$\pi D(1867)$-spectrum of Fig. \ref{fig:3}. Depending on the precise values of the parameters
the sextet state may be detected most efficiently via its coupling to the $\eta \,D(1867)$ channel
utilizing the $\eta - \pi_0$-mixing effect. We do not find a clear signal for
the sextet state in the $(1,1)$-sector reflecting the fact that the amount of
attraction for the sextet state is weaker in the scalar sector
as compared to the attraction in axial-vector sector.
However, a $(0,-1)$-bound state of mass 2353 MeV is predicted.

{\bfseries{Acknowledgments}}

M.F.M.L. acknowledges useful discussions with E.E. Kolomeitsev.


\begin{thebibliography}{9}

\bibitem{KL0307133}
E. E. Kolomeitsev and M.F.M. Lutz, GSI-Preprint-2003-20, hep-ph/0307133.

\bibitem{BaBar}

BABAR Collaboration, B. Aubert et al.,  Phys. Rev. Lett. {\bf 90} (2003) 242001.

\bibitem{CLEO}

CLEO Collaboration, D. Besson et al., hep-ex/0305017, hep-ex/0305100.

\bibitem{BELLE}

BELLE Collaboration, hep-ex/0307021.

\bibitem{NRZ93}

M.A. Nowak, M. Rho and I. Zahed, Phys. Rev. {\bf D 48} (1993) 4370

\bibitem{BH94}

W.A. Bardeen and C.T. Hill, Phys. Rev. {\bf D 49} (1994) 409.

\bibitem{BEH03}

W.A. Bardeen, E.J. Eichten and Ch.T. Hill, hep-ph/035049.

\bibitem{NRZ03}

M.A. Nowak, M. Rho and I. Zahed, hep-ph/0307102.

\bibitem{all-papers}

R.N. Cahn and J.D. Jackson, hep-ph/0305012;\\
T. Barnes, F.E. Close and H.J. Lipkin, hep-ph/0305025;\\
E. van Beveren and G. Rupp, hep-ph/0305035; help-ph/0306051;\\
H.-Y. Cheng and W.-S. Hou, hep-ph/0305038;\\
A.P. Szczepaniak, hep-ph/0305060;\\
S. Godfrey, hep-ph/0305122;\\
P. Colangelo and F. De Fazio, hep-ph/0305140;\\
G. S. Bali, hep-ph/0305209;\\
K. Terasaki, hep-ph/0305213;\\
S. Nussinov, hep-ph/0306187;\\
Y.-B. Dai, C.-S. Huang, C. Liu and S.-L. Zhu, hep-ph/0306274;\\
A. Dougall, R.D. Kenway, C.M. Maynard and C. McNelie, hep-lat/0307001;\\
Th.E. Browder, S. Pakvasa and A.A. Petrov, hep-ph/0307054;\\
A. Deandrea, G. Nardulli, A.D. Polosa, hep-ph/0307069;\\
Ch.-H. Chen and H.N. Li, hep-ph/0307075;\\
M. Sadzikowski, hep-ph/0307084;\\
A. Datta and P.J. O'donnell, hep-ph/0307106;\\
M. Suzuki, hep-ph/0307118.

\bibitem{IW91}
N. Isgur and M.B. Wise, Phys. Rev. Lett. {\bf 66} (1991) 1130.

\bibitem{KKP92}
U. Kilian, J.K. K\"orner and D. Pirjol, Phys. Lett. {\bf B 288} (1992) 360.

\bibitem{C95}
P. Colangelo et al., Phys. Rev. {\bf D 52} (1995) 6422.

\bibitem{Casalbuoni}
R. Casalbuoni et al., Phys. Rep. {\bf 281} (1997) 145.

\bibitem{LK00}
M.F.M. Lutz and E. E. Kolomeitsev, Proc. of Int. Workshop XXVIII
on Gross Properties of Nuclei and Nuclear Excitations, Hirschegg,
Austria, January 16-22, 2000.

\bibitem{LK01}
M.F.M. Lutz und E.E. Kolomeitsev, Found. Phys. {\bf 31} (2001) 1671.

\bibitem{LH02}
M.F.M. Lutz, GSI-Habil-2002-1.

\bibitem{LK02} M.F.M. Lutz and E. E. Kolomeitsev, Nucl. Phys. {\bf A
700} (2002) 193.

\bibitem{Granada}
C. Garc\'\i a-Recio, M.F.M. Lutz and J. Nieves, GSI-Preprint-2003-16, nucl-th/0305100.

\bibitem{Copenhagen}
E.E. Kolomeitsev and M.F.M. Lutz, GSI-Preprint-2003-17, nucl-th/0305101.

\bibitem{LK03}
M.F.M. Lutz and E. E. Kolomeitsev, GSI-Preprint-2003-19, nucl-th/0307039.

\bibitem{Rupp86}
E. Van Beveren et al., Z. Phys. {\bf C 30} (1986) 615.

\bibitem{WI90}

J.D. Weinstein and  N. Isgur, Phys. Rev. {\bf D 41} (1990) 2236.

\bibitem{JPHS95}
G. Janssen, B. C. Pearce, K. Holinde, J. Speth, Phys. Rev. {\bf D 52} (1995) 2690.

\bibitem{OOP99}
J.A. Oller, E. Oset and  J.R. Pelaez, Phys. Rev. {\bf D 59} (1999) 074001;
Erratum-ibid. {\bf D 60} (1999) 099906.

\bibitem{NVA02}

J. Nieves, M.P. Valderramaa and E.R. Arriola, Phys. Rev. {\bf D 65} (2002) 036002.

\bibitem{NP02}
A.G. Nicolaa and J.R. Pelaez, Phys. Rev. {\bf D 65} (2002) 054009.

\bibitem{Flyn-Nieves-91}
J.M. Flynn and J. Nieves, Phys. Lett. {\bf B 505} (2001) 82.

\bibitem{Wein-Tomo} S. Weinberg, Phys. Rev. Lett. {\bf 17} (1966) 616;
Y. Tomozawa, Nuov. Cim. {\bf A46} (1966) 707.

\bibitem{Wise92}

M.B. Wise, Phys. Rev. {\bf D 45} (1992) 2188.

\bibitem{YCCLLY92}
T.-M. Yan et al., Phys. Rev. {\bf 46} (1992) 1148.

\bibitem{BD92}
G. Burdman and J. Donoghue, Phys. Lett. {\bf 280} (1992) 287.

\bibitem{Jenkins94}
E. Jenkins, Nucl. Phys. {\bf B 412} (1994) 181.


\bibitem{PDG02}
K. Hagiwara et al., Phys. Rev. {\bf D 66} (2002) 010001.





\end{thebibliography}
\end{document}